\documentclass{webofc}

\usepackage[varg]{txfonts}   
\usepackage{hyperref}
\usepackage{url}
\usepackage{booktabs}
\usepackage{tabularx}
\usepackage{array}
\usepackage[dvipsnames]{xcolor}%
\usepackage{color}%
\usepackage[normalem]{ulem}
\hypersetup{colorlinks=true,citecolor=blue,urlcolor=blue,linkcolor=blue}
\newif\ifshowchanges

\showchangestrue

\colorlet{revone}{Green!80!black}   
\colorlet{revtwo}{Cyan!80!blue}     

\ifshowchanges
  \DeclareRobustCommand{\del}[1]{\textcolor{red}{\sout{#1}}}

\else
  \DeclareRobustCommand{\del}[1]{}

\fi
\begin{document}
\title{Review of LHD-ECRH activities and beyond}
%
%

\author{\firstname{Masaki} \lastname{Nishiura}\inst{1,2}\fnsep\thanks{\email{nishiura@nifs.ac.jp}} \and
        \firstname{Hiroe} \lastname{Igami}\inst{1} \and
        \firstname{Hiromi} \lastname{Takahashi}\inst{1} \and
        \firstname{Naoki} \lastname{Kenmochi}\inst{1} \and
        \firstname{Ryoma} \lastname{Yanai}\inst{1} \and
        \firstname{Kenji} \lastname{Ueda}\inst{1} \and
        \firstname{Yoshinori} \lastname{Mizuno}\inst{1} \and
        \firstname{Toshiki} \lastname{Takeuchi}\inst{1} \and
        \firstname{Yasuo} \lastname{Yoshimura}\inst{3} \and
        \firstname{Takashi} \lastname{Shimozuma}\inst{1} \and
        \firstname{Shin} \lastname{Kubo}\inst{4} \and
        \firstname{Toru} \lastname{Ii Tsujimura}\inst{5} \and
        \firstname{Sakuji} \lastname{Kobayashi}\inst{5} \and
        \firstname{Tsuyoshi} \lastname{Kariya}\inst{6}
}

\institute{National Institute for Fusion Science, 322-6 Oroshi, Toki, Gifu 509-5292, Japan
\and
           Graduate School of Frontier Sciences, The University of Tokyo, Kashiwanoha, Chiba 277-8561, Japan 
\and
           National Institutes for Quantum Science and Technology, Naka, Ibaraki 311-0193, Japan
\and
           Chubu University, Kasugai, Aichi 487-8501, Japan
\and
           Kyoto Fusioneering Ltd., Ota-ku, Tokyo 143-0006, Japan
\and
           Plasma Research Center, University of Tsukuba, Ibaraki 305-8577, Japan
}

\abstract{The electron cyclotron resonance heating (ECRH) system on the Large Helical Device (LHD) played a key role in the progress of helical fusion research from the first plasma in 1998 to the final discharge on December 25, 2025. This paper presents a historical review of ECRH activities on LHD, highlighting the stepwise advances in hardware, including gyrotrons, transmission lines, and launcher systems, together with the exploration of advanced plasma regimes such as overdense heating and long-pulse steady-state operation. Beyond plasma heating, the ECRH system was also used as a high-power probe for advanced diagnostics, including collective Thomson scattering (CTS) as well as correlation electron cyclotron emission (CECE), and as a flexible actuator for transport studies and real-time control experiments. Finally, we discuss how the technological assets and operational expertise accumulated over the nearly three decades of the LHD program are being further developed and applied to next-generation helical devices, particularly CHD and CHD-U, as well as to broader gyrotron and millimeter-wave applications.
}
\maketitle
\section{Introduction and Role of ECRH in LHD}
The Large Helical Device (LHD) is a superconducting heliotron with helical mode numbers of $l/m=2/10$, a major radius of $R_0=3.9$ m, a minor radius of $a=0.63$ m, a plasma volume of approximately $30~\mathrm{m}^3$, and a maximum magnetic field of 3 T at the magnetic axis. LHD began operation on March 31, 1998, and completed its final discharge on December 25, 2025. The principal auxiliary heating systems on LHD were neutral beam injection (NBI), ion cyclotron range of frequency (ICRF) heating, and electron cyclotron resonance heating (ECRH). In the final stage of LHD operation, the operational ECRH system consisted of three gyrotrons: one 77 GHz gyrotron and two 154 GHz gyrotrons, with a total nominal short-pulse power of approximately 3 MW. The number of operational gyrotrons had been reduced because of gyrotron failures and the relocation of power-supply equipment. ECRH played several distinct roles in LHD experiments. In addition to reliable plasma start-up, it enabled high electron temperatures, long-pulse plasma sustainment, localized transport studies, and active control of plasma parameters. 

During the early phase of LHD operation, ECRH demonstrated central electron temperatures exceeding 10 keV, electron heating in high-density plasmas, and a 756 s discharge sustained by continuous-wave (CW) ECRH \cite{Kubo_2005}. The development and operational status of the LHD-ECRH system during its first decade, including the evolution of gyrotrons, transmission lines, launcher systems, and steady-state capability, were comprehensively reviewed by Shimozuma et al. \cite{Shimozuma_2010}. However, long-pulse operation at this stage was limited to low-density, relatively low-temperature plasmas and was constrained by engineering limitations in the transmission system.

The development of these capabilities required not only higher gyrotron output power but also improvements in millimeter-wave transmission, steerable launcher systems, cooling, polarization control, and quantitative analysis of power deposition. Precise control of the deposition location and injected power, together with rapid power modulation, also enabled ECRH to provide well-defined localized perturbations to the plasma. These capabilities extended the application of the LHD-ECRH system beyond plasma heating to diagnostics, transport studies, and active control experiments. This paper provides a concise historical overview of the development and major achievements of the LHD-ECRH system presented at EC-23, and discusses how the technologies and operational concepts developed on LHD will be carried forward into the next-generation CHD and CHD-U projects at NIFS.

\section{Progress of the LHD-ECRH system} 

\subsection{Evolution of gyrotrons} 
\begin{figure}[ht]
    \centering
    \includegraphics[width=0.95\linewidth]{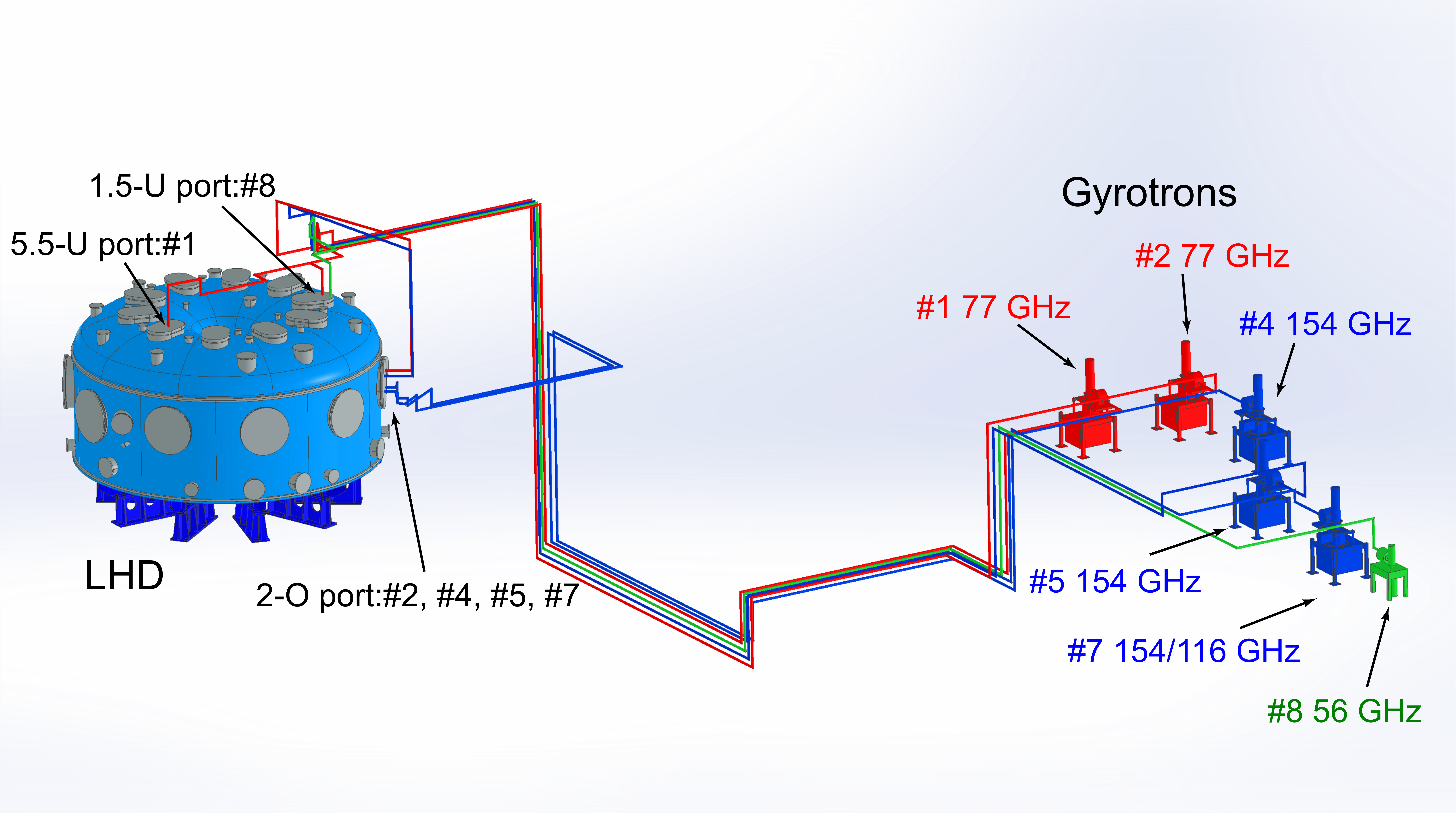}
    \caption{Schematic layout of the gyrotrons and transmission lines connecting them to the LHD-ECRH injection ports during 2021--2024. Adapted from R. Yanai et al.~\cite{Yanai_2026}, with minor changes to the figure labels. }
    \label{fig:LHD-gyrotrons}
\end{figure}

\begin{figure}[ht]
    \centering
    \includegraphics[width=0.95\linewidth]{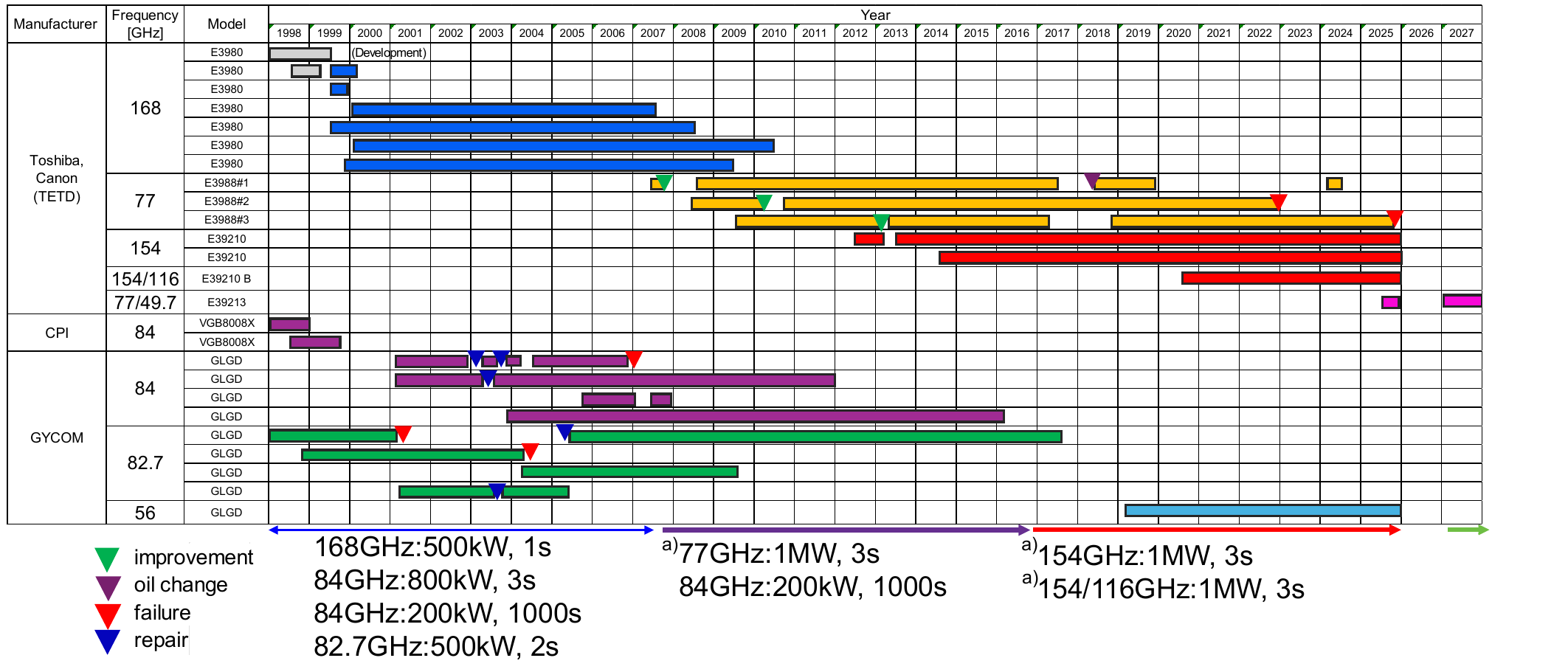}
    \caption{Historical timeline of the implementation and operation of gyrotrons on LHD, including the causes of operational interruption up to 2025. The gyrotron selected for transfer to CHD will be operated from 2027. This figure is adapted and updated from H. Takahashi et al. \cite{Takahashi_2026}. $^\mathrm{a)}$ These gyrotrons were operated at a few hundred kW in steady-state operation.}
    \label{fig:gyrotrons}
\end{figure}

Figure~\ref{fig:LHD-gyrotrons} shows the arrangement of the LHD gyrotrons, transmission lines, and principal ECRH injection ports discussed in this section. The major gyrotrons used on LHD are summarized in Table~\ref{tab:gyrotrons}. The evolution is shown in figure~\ref{fig:gyrotrons}. The early evolution of the LHD-ECRH system, including the successive introduction of 84, 168, and 77 GHz gyrotrons and the associated upgrades of transmission and launcher systems, was reviewed in detail by Shimozuma et al. \cite{Shimozuma_2010}. The present review focuses on how these successive system developments expanded the accessible plasma regimes and broadened the applications of ECRH in LHD experiments. In the initial stage, 84 GHz gyrotrons were used mainly for plasma start-up and fundamental-resonance heating, together with a 168 GHz gyrotron for second-harmonic heating. As the ECRH system was upgraded, the earlier gyrotrons were progressively replaced by megawatt-class 77 and 154 GHz gyrotrons. These later gyrotrons provided not only higher output power but also improved long-pulse capability, efficiency, reliability, and operational flexibility \cite{Kariya_2017}. Megawatt-class gyrotrons were developed through collaborations with gyrotron-development groups and manufacturers, while the LHD ECRH group integrated these sources into the heating system and optimized their operation for LHD experiments. The overall history of their implementation and operation on LHD has recently been summarized by Takahashi et al. \cite{Takahashi_2026}. A dual-frequency 154/116 GHz gyrotron was developed to extend the accessible magnetic-field range, and a 77/49.7 GHz dual-frequency gyrotron was introduced in the final stage of LHD operation. A 56 GHz gyrotron was also used for plasma start-up at low magnetic field during deuterium experiments, where experience showed that reliable plasma initiation using NBI alone was difficult. These successive upgrades expanded the usable magnetic-field range, improved the availability and reliability of high-power ECRH, and allowed the heating configuration to be adapted more flexibly to different experimental scenarios. Thus, the evolution of the gyrotron system involved not simply an increase in output power, but also substantial improvements in frequency flexibility, engineering maturity and operational usability.

\begin{table}[htbp] 
\centering \caption{Nominal specification of gyrotrons used in LHD.} \label{tab:gyrotrons} \small \setlength{\tabcolsep}{4pt} \renewcommand{\arraystretch}{1.05} 
\begin{tabularx}{\textwidth}{ >{\centering\arraybackslash}p{1.35cm} >{\centering\arraybackslash}p{1.55cm} >{\centering\arraybackslash}p{1.9cm} >{\centering\arraybackslash}p{3.25cm} >{\raggedright\arraybackslash}X } 
\toprule Frequency [GHz] & Model & Mode & Power / Pulse & Primary role / Period \\ 
\midrule 168 & E3980 & $\mathrm{TE}_{31,8}$ & 500 kW / 1 s & Second-harmonic heating \\ 
84 & VGB8008X & $\mathrm{TE}_{15,2}$ & 800 kW / 3 s & Fundamental heating \\ 
84 & GLGD & $\mathrm{TE}_{15,2}$ & 200 kW / 1000 s & Long-pulse operation \\ 
82.7 & GLGD & $\mathrm{TE}_{12,6}$ & 500 kW / 2 s & Fundamental heating \\ 
77 & E3988 & $\mathrm{TE}_{18,6}$ & $\sim$1 MW / several s$^{\mathrm{a)}}$ & Principal ECRH source \\ 
154 & E39210 & $\mathrm{TE}_{28,8}$ & $\sim$1 MW / several s$^{\mathrm{a)}}$ & High-density heating \\ 
154/116 & E39210B & $\mathrm{TE}_{28,9}$ / $\mathrm{TE}_{21,7}$ & $\sim$1 MW / several s$^{\mathrm{a)}}$ & Dual-frequency operation \\ 
56 & GLGD & $\mathrm{TE}_{8,3}$ & 400 kW / 1 s & Low-field start-up \\ 
77/49.7 & E39213 & $\mathrm{TE}_{18,6}$ / $\mathrm{TE}_{12,4}$ & $\sim$1 MW / several s$^{\mathrm{a)}}$ & 77 GHz used in LHD \\ 
\bottomrule \end{tabularx} 
\raggedright {\footnotesize $^{\mathrm{a)}}$ A few hundred kilowatts in CW operation.}
\end{table} 

\subsection{Transmission line, launcher, and power-deposition analysis} 
Progress in the gyrotron development was accompanied by continuous improvements in the transmission and launching systems. Evacuated corrugated waveguides with inner diameters of 3.5 inches (88.9 mm) and 1.25 inches (31.75 mm) were employed for long-distance transmission from the gyrotron room to the LHD torus hall. Water cooling was introduced to improve reliability during high-power and long-pulse operation, while polarization-control components enabled flexible control of the injected EC-wave polarization. The importance of evacuating and water-cooling the transmission lines, reducing arcing, and improving transmission efficiency for high-power and steady-state operation was already recognized during the first decade of LHD-ECRH development \cite{Shimozuma_2010}.

At the 2-O equatorial port, four independent ECRH launchers, 2-OLR, 2-OLL, 2-OUR, and 2-OUL, were installed inside the vacuum vessel. Their in-vessel mirrors were water-cooled, and their final mirrors were steerable in both the toroidal and poloidal directions using two-axis drive systems with ultrasonic motors or servomotors operated from the air side. This flexibility enabled control of the EC wave power-deposition location through on-axis or off-axis heating, supporting both heating and transport studies. The steerable launchers were also used in plasma diagnostic applications requiring spatially resolved measurements. 

Quantitative analysis of ECRH power deposition also became an integral part of LHD operation. ECRH injection data were incorporated into the LHD AutoAna framework, and the LHDGauss \cite{Tsujimura_2015} and LHDGauss2 \cite{Yanai_2023} ray-tracing codes were used to calculate EC-wave trajectories and power-deposition profiles based on plasma equilibrium and measured electron-density and temperature profiles. These deposition data were routinely used for time-resolved heat-transport analysis. 

The combination of source capability, robust transmission, flexible launching, and quantitative deposition analysis was essential for the system-level maturity of LHD-ECRH. 

\section{Plasma regimes enabled by ECRH} 

\subsection{Long-pulse and high-temperature operation} 
Long-pulse operation was one of the major objectives of LHD. In an early ECRH-only experiment, an 84 GHz CW gyrotron sustained a low-density plasma for 756 s with approximately 72 kW of injected ECRH power. The average electron density was about $2.4\times10^{17}~\mathrm{m}^{-3}$, and the ECE radiation temperature was approximately 240 eV \cite{Kubo_2005}. Although the plasma parameters were still modest, this experiment demonstrated the feasibility of long-pulse ECRH operation and identified important engineering limitations. The discharge was eventually terminated by a pressure increase in the ECRH transmission system caused by outgassing associated with the temperature rise of the waveguide components. Subsequent reassessment and water cooling of the transmission components enabled a much longer discharge of 3900 s with approximately 110 kW of ECRH power \cite{Shimozuma_2010}.

Continuous improvements in gyrotron reliability, water-cooled transmission, plasma-wall conditioning, and power-deposition control subsequently enabled substantially higher-performance plasmas to be sustained for long durations. In 2015, a helium plasma with an average electron density of approximately $n_{\mathrm{e,ave}}=1.1\times10^{19}~\mathrm{m}^{-3}$, a central electron temperature of $T_{\mathrm{e}0}\sim2.5$ keV, and a central ion temperature of $T_{\mathrm{i}0}\sim1.0$ keV was sustained for 39 minutes using a dual-frequency ECRH scenario, in which two 154~GHz gyrotrons were operated continuously and two 77~GHz gyrotrons were operated alternately at 2-minute intervals to mitigate their internal pressure rise \cite{Yoshimura_2016}. The discharge was eventually terminated by an impurity influx.

Subsequently, stable sustainment of an electron internal transport barrier (e-ITB) was demonstrated for more than 5 min using ECRH alone. A helium plasma with $n_{\mathrm{e,ave}}=1.1\times10^{19}~\mathrm{m}^{-3}$ and $T_{\mathrm{e}0}\sim3.5$ keV was maintained with approximately 340 kW of ECRH power, demonstrating that localized on-axis ECRH could sustain an improved-confinement state over long pulse durations \cite{Yoshimura_2018}.

In the present historical perspective, the extension of the accessible density range is attributed to the introduction of 154 GHz ECRH, which reduced beam refraction and provided a higher cutoff density, together with optimization of the EC-wave injection geometry. At shorter pulse durations, the increase in available ECRH power substantially expanded the accessible high-$T_{\mathrm{e}}$ regime over a broad range of plasma densities, with central electron temperatures exceeding 10 keV \cite{Takahashi_2017}. These results demonstrated that the improvement in plasma performance was achieved through the development and integration of the entire ECRH system, rather than simply by increasing the output power of individual gyrotrons. 

\subsection{Heating beyond conventional cutoff limits} 
LHD also explored ECRH scenarios that extended the accessible plasma parameter range beyond that of the conventional fundamental and second-harmonic heating schemes. For typical operation around $B_{\mathrm{t}}\simeq2.75$ T, fundamental 77 GHz O-mode and second-harmonic 154 GHz X-mode waves were primarily used because of their high absorption efficiency. To extend the accessible density and magnetic-field ranges, both high-harmonic heating and mode-conversion schemes involving electron Bernstein waves (EBWs) were investigated.

High-harmonic ECRH using second-harmonic O-mode (O2) and third-harmonic X-mode (X3) waves was investigated by optimizing the injection geometry and magnetic-field configuration. With 77 GHz O2 injection, an absorption rate of approximately 30--40\% was maintained even above the cutoff density of the conventional X2 mode. For X3 heating, a maximum absorption rate of approximately 40\% was obtained around $n_{\mathrm{e}}\sim1.5\times10^{19}~\mathrm{m}^{-3}$ and $T_{\mathrm{e}}\sim1.2$ keV. Stepwise injection from three gyrotrons with a total power of approximately 3 MW increased the central electron temperature from about 0.6 keV to 2.2 keV, demonstrating that the elevated electron temperature could enhance the absorption of subsequent high-harmonic EC-wave injection \cite{Shimozuma_2015}.

In the O--X--B scheme, an O-mode wave launched from the low-field side is converted to an X-mode near the O-mode cutoff layer and subsequently to an EBW near the upper-hybrid resonance (UHR) layer. In LHD, the O--X conversion window is particularly narrow because of the relatively large density scale length. Nevertheless, an increase in plasma stored energy was experimentally demonstrated when the launch direction was optimized for O--X--B mode conversion \cite{Igami_2012}. 

High-field-side slow X--B heating was also demonstrated. An X-mode millimeter-wave beam was redirected from the high-field side by an in-vessel mirror, and converted to an EBW near the UHR layer. An increase in electron temperature was observed in a density regime above the conventional cutoff limit \cite{Yoshimura_2013}. 

These experiments demonstrated that the flexibility of the LHD launcher system could be exploited not only to optimize conventional ECRH but also to access high-density plasma regimes through alternative propagation and mode-conversion schemes. 

\subsection{Development of alternative injection geometry} 

During the later stage of the LHD project, alternative injection geometries were investigated to improve the accessibility of EC waves to the central resonance region. In high-density plasmas, strong refraction during standard oblique injection can distort the beam trajectory and degrade beam focusing, making it difficult to maintain a well-focused beam directed toward the central resonance layer. In addition, oblique injection introduces a Doppler shift of the EC resonance, which can displace the effective heating location from the magnetic axis. A nearly perpendicular injection geometry from the 1.5-UO port was therefore investigated. 

Ray-tracing calculations showed that injection normal to a magnetic flux surface, i.e. \del{a} perpendicular injection, provided a substantially straighter propagation path than standard oblique injection and reduced the Doppler-shift-induced displacement of the resonance location. Experiments using 1.5-UO injection demonstrated clear electron heating together with an increase in plasma stored energy \cite{Yanai_2026, Nishiura_2022}. These results showed that optimization of the injection geometry can mitigate refraction effects, improve access to the intended central resonance region, and extend the applicability of ECRH beyond the constraints associated with standard oblique injection.

\section{Beyond heating: diagnostics and extended applications of ECRH} 

\subsection{Collective Thomson scattering} 
The millimeter-wave technology developed for ECRH enabled advanced plasma diagnostics. In particular, a high-power 77 GHz ECRH beam was used as the probe beam for collective Thomson scattering (CTS). The CTS receiver system shared part of the millimeter-wave transmission infrastructure developed for ECRH. 

CTS measures collective electron-density fluctuations associated with ion motion and provides information on ion temperature and ion velocity distributions, including energetic-ion populations. On LHD, CTS was applied to measurements of energetic ions produced by tangential neutral beam injectors with beam energies of approximately 180 keV \cite{Nishiura_2014}. 

Comparison of the measured CTS spectra with forward-modeling simulations showed that the observed spectral asymmetry could be explained by anisotropic energetic-ion velocity-space distributions.  The CTS receiver system was also extended to correlation electron cyclotron emission (CECE) measurements, adding further diagnostic capability for electron-temperature fluctuation and turbulence studies. These applications extended the use of ECRH-related millimeter-wave technology from plasma heating to velocity-space and turbulence diagnostics.

\subsection{ECRH as a controlled perturbation for transport studies} 
The excellent localization and modulation capability of ECRH, which can produce an electron internal transport barrier, made it useful as an active perturbation source for transport studies. In the later stage of the LHD project, modulated on-axis ECRH was used as a controlled perturbation, while a heavy ion beam probe (HIBP) measured the resulting temporal response of the plasma potential.

The experiments revealed a clear causal relationship between the local electron-temperature response and the plasma-potential response. Following modulation of the ECRH power, the electron temperature near the deposition region responded first, followed by the local plasma potential on a millisecond timescale. In contrast, outside the e-ITB region, no significant delay was observed between the electron-temperature and plasma-potential responses.

The measurements also allowed the plasma-potential dynamics during the formation and decay of electron internal transport barriers to be compared between hydrogen and deuterium plasmas, revealing no significant isotope dependence in the potential response. These results demonstrated the use of ECRH as a well-controlled local actuator for investigating causal relationships among heating, electron temperature, plasma potential, and transport.

The improved HIBP capability further enabled plasma-potential measurements under a wider range of ECRH conditions \cite{Nishiura_2025}. Together with recent LHD experiments involving anisotropic energetic-ion control, these measurements provide opportunities for investigating possible relationships among energetic-ion distributions, plasma potential, and transport.

\subsection{Real-time control applications}
The flexibility of the LHD-ECRH system was also utilized in real-time control experiments. FPGA-based control systems were developed to adjust the EC-wave polarization and deposition location in response to changes in plasma conditions. Fast steering of the launcher mirrors was also applied to optimize beam alignment for CTS measurements. More recently, data-driven approaches were investigated for generating ECRH control parameters and for radiative-collapse avoidance. These developments and their applications are reviewed in detail by Kenmochi et al. \cite{Kenmochi_2026}. 

\section{ECRH system plans for CHD and CHD-U}

\begin{figure}[!b]
    \centering
    \includegraphics[width=0.6\linewidth]{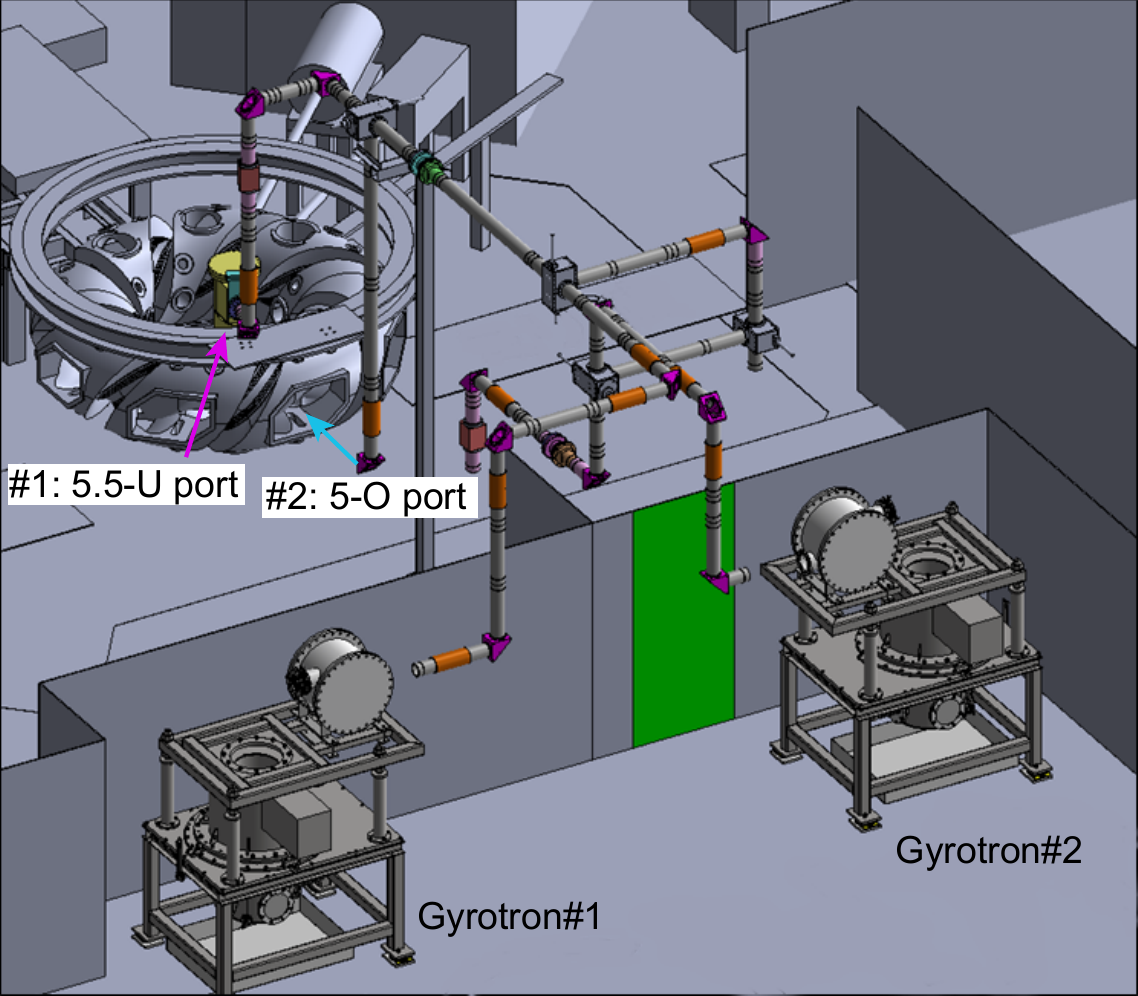}
    \caption{Layout of the CHD ECRH system and gyrotron area. Initially, Gyrotron \#1 will be used for switchable injection through the 5.5-U and 5-O ports. After March 2028, Gyrotron \#2 is planned to be connected to the 5-O port, enabling two-gyrotron ECRH operation. A dummy load (not shown) is shared between the two transmission lines via waveguide switches.}
    \label{fig:CHD-ECRH}
\end{figure}

The initial CHD ECRH system is planned to reuse selected equipment from LHD: one E39213\#1 dual-frequency 77/49.7 GHz gyrotron, two anode power supplies, two body power supplies, and one cathode power supply. This transfer will provide a starting point for commissioning the CHD ECRH system while carrying forward equipment and operational experience from LHD. In the first stage, the gyrotron transmission line will be switchable between the 5-O and 5.5-U injection ports, as shown in Fig.~\ref{fig:CHD-ECRH}. A second gyrotron line is planned for a later stage. With both lines in operation, the total injected power is expected to be approximately 2 MW.

For CHD-U, the current target is to develop the power-supply and gyrotron infrastructure to support operation of up to seven gyrotrons and ECRH injection of up to 5 MW by March 2030. Dual-frequency 77/49.7 GHz gyrotrons are planned. The gyrotron configuration for achieving 5 MW has not yet been selected: options under consideration include operating four 1.5 MW gyrotrons or five 1 MW gyrotrons. These are development plans, and the final configuration and schedule will depend on detailed design and procurement.

\section{Lessons learned and legacy for CHD/CHD-U and beyond}

The experience accumulated during nearly three decades of LHD operation demonstrated that high gyrotron output power alone is insufficient to realize the full potential of ECRH as a tool for advanced plasma research. Its scientific capability relies on the integration of several key elements:

\begin{enumerate}
\item high-power and, where required, long-pulse gyrotron capability;
\item robust and well-cooled millimeter-wave transmission;
\item flexible launcher systems capable of controlling the relative geometry between the plasma and the injected EC beam;
\item quantitative calculation and validation of power deposition;
\item sufficient heat-load handling capability of in-vessel components for high-power microwave injection; and
\item integration with plasma diagnostics and active perturbation experiments.
\end{enumerate}

Together, these developments substantially broadened both the operational capabilities and scientific applications of LHD-ECRH. Long-pulse and high-temperature operation, overdense heating, millimeter-wave diagnostics using CTS and CECE, and controlled perturbation experiments are representative outcomes of this system-level development. The most important legacy of LHD-ECRH is therefore not only its hardware, but also the integrated approach to ECRH system development and operation established through the LHD program. This approach will be carried forward to CHD and CHD-U, where gyrotrons and power supplies will be integrated with transmission, launching, power-deposition analysis, and heat-load handling capabilities. The experience gained on LHD will support the commissioning and operation of CHD, whose first plasma is planned for March 2027, and the subsequent development of CHD-U.

The LHD-ECRH system is planned to be renovated and reorganized as the ECRH system for CHD-U, rather than simply being decommissioned following the completion of LHD operation. Some of the gyrotrons and associated facilities are already being utilized for gyrotron development in collaboration with fusion start-up companies, industrial applications, and academic research. In collaboration with UKAEA and other partners, the expertise and experimental infrastructure established through LHD-ECRH activities have also been utilized for the development of a 28/35 GHz dual-frequency gyrotron for MAST-U and related studies toward STEP. Toward CHD-U operation, an operational framework will be established for the effective and flexible use of up to seven gyrotrons for these diverse activities as well as for plasma heating experiments. Thus, the LHD-ECRH infrastructure will continue to serve not only as the technological basis for the CHD-U ECRH system, but also as a platform for gyrotron development, international collaboration, millimeter-wave applications, and broader academic and industrial research.

\section*{Acknowledgments} 
The authors sincerely thank all researchers, engineers, and technical staff who contributed to LHD-ECRH activities over nearly three decades. In particular, we thank Mr. Itoh, Mr. Okada, and Mr. Takita for their dedicated engineering support, which made the long-term operation and continuous improvement of the ECRH system possible.

%
%
%

\end{document}